\documentclass[11pt]{article}
\begin{document}
\baselineskip .3in

 {\Large{\bf Fine Structure Constant and Fractal Potential}}

 \vskip .2in
 {\bf { A. Bhattacharya $^{1}$, R. Saha and R. Ghosh }}

Department of Physics, Jadavpur University, Calcutta 700032,
India.

\vskip 0.001in

e-mail:$^{1}$pampa@phys.jdvu.ac.in

 \vskip .003in

\vskip 0.01in

{\small{ {\bf Abstract}
\vskip .02in

The variation of fine structure constant with the variation of
velocity of light has been studied considering contribution to the
permeability and permittivity of the vacuum and incorporating
contribution of fractal potential originated from the
non-differentiability of fractal fluctuation. The time variation
of $\alpha$ is found to depend on the complex velocity field
characteristics of fractal nature of space time.

\vskip .05in

 }}
\vskip .1in
{\bf Introduction:}

Fine structure constant represents interaction between two charged
particle and is expressed as $\alpha =
\frac{e^{2}}{4\pi\epsilon_{0}c\hbar}$. This coupling constant is
responsible for the stable structure of atoms and is supposed to
be a fundamental constant of nature. Recent experimental
observations hint at the fact that the fine structure constant may
not be constant [1] and varies according to the 'large number'
hypothesis by Dirac [2]. QED suggests that the fine structure
constant is not a constant rather it varies on distance and energy
[3,4]. Some experimental results indicate the variation of
$\alpha$ from cosmological time to the present. One of the
important candidate for variation of $\alpha$ is the variable
speed of light(c). In cosmology the variable speed of light is
required to solve a number of cosmological problems like flatness.
In QED, vacuum polarization plays an important role for the
concept of variable $\alpha$. Haranas [5] has investigated the
variation of $\alpha$ with the change of speed of light. They have
argued that the variation of $\alpha$ may be attributed to the
variable speed of light from cosmological time to the present
value. Considering c as function of time c(t), they have obtained
$\mid\frac{\alpha'}{\alpha}\mid = 10^{-15} - 10^{-16} yr^{-1}$ at
z=3. Tobar [6] has argued that the variation of $\alpha$ is
manifested due to the differential change in the fraction of
quanta of the electric and magnetic flux of force of electron
whereas variation of $\hbar c$ has shown to manifest due to common
mode change. Leefer et al [7] have investigated the variation
of fine structure constant using atomic Dysprosium. The transition
between the nearly degenerate excited states with opposite parity
is found to depend on $\alpha$. Ranada [8] has proposed a model
to explain the observed cosmological variation of fine structure
constant as an effect of quantum vacuum. He has pointed out that
due to the fourth Heisenberg relation, the density of the sea of
virtual particles in quantum vacuum must change in a gravitational
field with the corresponding change in permittivity and
permeability. The variation of fine structure constant and its
relation with fractal space has been studied by Bhattacharya et al
[9]. They have investigated the effect of space-time geometry
to the variation of $'\alpha'$ between matter and radiation
dominated era and have suggested that the variation of $'\alpha'$
may be attributed to the intrinsic scale dependence of fundamental
constants. Tedesco [10] has studied the variation of fine
structure constant due to generalised uncertainty principle in the
space-time domain wall. They have pointed out that the fine
structure constant may be different in different epoch considering
the generalized Heisenberg uncertainty principle which include
gravitational interaction. Moffat [11] has investigated the
variation of fine structure constant with speed of light and
observed that the varying speed of light reproduces the
observational data well in the range.

\vskip.001in

In the present work we have investigated the variation of fine
structure constant considering the contribution of fractal
potential arising due to the fractal nature of vacuum. We have
studied the time variation of fine structure constant and have
observed that $\alpha$ varies with derivative of $\Delta \nu$
where $\nu$ is the velocity field.

\vskip.001in
{\bf Formalism:}
\vskip.01in

The fine structure constant in vacuum can be expressed as:

\begin{equation}
\alpha = \frac{e^{2}}{4\pi c \epsilon_{0}}
 \end{equation}

Now considering the effect of medium we can rewrite the above
expression as [8]:
\begin{equation}
\alpha' = \alpha \sqrt{\frac{\mu_{r}}{\epsilon_{r}^{3}}} = \alpha
\frac{\sqrt{\epsilon_{r}\mu_{r}}}{\epsilon_{r}^{2}}
\end{equation}

where $e$ is $\frac{e}{\epsilon_{r}}$ and $ c =
\frac{c}{\sqrt{\epsilon_{r}\mu_{r}}}$ in medium. It may be mentioned that
the permittivity and permeability of the quantum vacuum changes to
$\epsilon_{r}\epsilon_{0}$ and $\mu_{r}\mu_{0}$ indicating the
effect of thickening and lightening of the medium. Recently Ranada
[8] has suggested a model to investigate the variation of
$\alpha$ where quantum vacuum has been suggested to be a
transparent optical medium characterized by its permittivity and
permeability and the change of $'\alpha'$ is suggested to be a
consequence of fourth Heisenberg relation applied to the
Gravitational interaction of virtual pairs. In such system the
observed quantities like permittivity and permeability of quantum
vacuum have been expressed as $\epsilon_{r}\epsilon_{0}$ and
$\mu_{r}\mu_{0}$ respectively. Incorporating the effect of
gravitational field $\phi_{0}$ along with Newtonian potential
$(\phi)$, the value of $\epsilon_{r}$ and $\mu_{r}$ at a space
time point $\phi$ have been expressed as:[8]

\begin{equation}
\epsilon_{r} = 1 - \frac{\beta(\phi-\phi_{0})}{c^{2}}
\end{equation}

\begin{equation}
\mu_{r} = 1 - \frac{\gamma(\phi-\phi_{0})}{c^{2}}
\end{equation}

where $\beta, \gamma$ are certain positive co-efficient, $\phi$ is
potential at a space-time point and $\phi_{0}$ is present
gravitational potential of all universe.

\vskip .001in

In scale relativity the space-time is described as
non-differential continuum [12]. We have introduce in effect of
fractal potential due to space time non-differentiability or
fractality. The fractal fluctuation can be expressed as effective
force of the potential energy from which it has been derived.
Nottale [12] has pointed out that in the scale relativistic
fractal space (time) approach, the force is manifestation of the
very structure of the space. The proposition applied to the
standard quantum mechanical case of microphysics, requires such
force to be universal and independent of the mass of particles.
It may be mentioned that the most interesting consequence of
non-differentiability of space time is the introduction of two
valuedness in velocity. Nottale [12] has pointed out that the
breaking of discrete symmetry like reflection in variance on the
differential element of time implies two valuedness of velocity
which in turn results in the origin of complex velocity and they
are fractal function of the resolution such as:[12]

\begin{equation}
V_{+}[x(t,dt),t,dt] = v_{+}[x(t),t] + w_{+}[x(t,dt),t,dt]
\end{equation}

\begin{equation}
V_{-}[x(t,dt),t,dt] = v_{-}[x(t),t] + w_{-}[x(t,dt),t,dt]
\end{equation}

where the fractal velocity field has been decomposed in terms of
their classical and fractal parts. The fractal fluctuation can be
expressed in terms of a fractal force and the potential energy
which is obtained from its derivative. In scale relativity and in
fractal space time approach is termed as manifestation of the
structure of the space. In such understanding as pointed out by
Nottale [12], the force then must be independent of mass of the
particle. The fractal potential energy can be expressed as
[12],

\begin{equation}
\phi_{F} = -imD \cdot \bigtriangledown \nu
\end{equation}

where $D = \frac{\hbar}{2m}$ and $\nu$ is complex velocity field
expressed as $\nu = V - iU$ where $V = \frac{v_{+}+v_{-}}{2}$, $U
= \frac{v_{+}-v_{-}}{2}$.

Including the contribution to potential from fractal space, ie.
fractal potential, the expression (3) and (4) can be recast as:

\begin{equation}
\epsilon_{r} = 1 - \frac{\beta( \phi - \phi_{0} - imD \cdot
\bigtriangledown \nu)}{c^{2}}
\end{equation}

\begin{equation}
\mu_{r} = 1 - \frac{\gamma( \phi - \phi_{0} - imD \cdot
\bigtriangledown \nu)}{c^{2}}
\end{equation}

so that,

\begin{equation}
\sqrt{\epsilon_{r}\mu_{r}} \simeq [1 -
\frac{(\beta+\gamma)(X-Y)}{2c^{2}}]
\end{equation}

where $X = \phi - \phi_{0}$, $Y = \frac{-imD \cdot
\bigtriangledown \nu}{c^{2}}$, we have

\begin{equation}
\mu_{r}^{2} = [1 - \frac{\beta(X-Y)}{c^{2}}]^{2}
\end{equation}

from (13),(14) and (2) we arrive at:

\begin{equation}
\frac{\alpha'}{\alpha} \simeq [1 +
\frac{Z(3\beta-\gamma)}{2c^{2}}]
\end{equation}

where Z = X-Y. With $\alpha' = \alpha + \bigtriangleup \alpha$,
(15) can be rewritten as:

\begin{equation}
\frac{\alpha+\bigtriangleup \alpha}{\alpha} = [1 +
\frac{Z(3\beta-\gamma)}{2c^{2}}]
\end{equation}

so that:
\begin{equation}
\frac{\bigtriangleup \alpha}{\alpha} = \xi(\phi - \phi_{0} - imD
\cdot \bigtriangledown \nu)
\end{equation}
 where $\xi = \frac{(3\beta-\gamma)}{2c^{2}}$.

  The variation of
 $'\alpha'$ with time is obtained as:

 $\hskip 1.25in \frac{1}{\alpha}\cdot\frac{d(\triangle \alpha)}{dt} = \xi \frac{d}{dt}(-imD\cdot\bigtriangledown \nu)$

\begin{equation}
= - \xi imD\cdot \frac{d\nabla \nu}{dt}
\end{equation}

where $'\nu'$ is the velocity field as stated earlier.

with $\frac{d}{dt}\nu = iD \triangle \nu$ from [12] we come
across:

\begin{equation}
\frac{1}{\alpha}\cdot\frac{d(\triangle \alpha)}{dt} = AmD^{2}
\nabla \cdot\triangle\nu
\end{equation}

The rate of change of $\alpha$ depends upon the fractal potential
part where $'\nu'$ is the complex velocity field.
\vskip.01in
{\bf Discussion}

In the present work we have investigated the  variation of fine
structure constant with variation of velocity of light resulted
from the variation of permittivity and permeability due to fractal
fluctuation of space-time. The time variation of $'\alpha'$ is
found to depend on the derivative of complex velocity field. The
contribution to the potential which results from the fractal
fluctuation has been incorporated to the vacuum potential. It may
be mentioned that Chang et al [13] have investigated the
observed Quasar absorption spectra and suggested that they can be
interpreted as the effect of space time inhomogeneity whereas
Timotte et al [14] have investigated the effect of fractal
potential in the system dynamic using fractal character of
particle movement. They have pointed out that complex speed field
contributed to establish coherence in the system which results
type I superconductivity. Recently Naschie [15] has
investigated cantorian space-time and high energy particle
physics. He has argued that recent results in particle physics
suggested the geometry of space-time to be a non-differential
contour set rather than flat and smooth. The radiative correction
due to vacuum polarization also impose some correction to the fine
structure constant substantially below compton wave length of
electron. Contribution of fractal fluctuation potential to the
vacuum potential is very important consequence in variation of
fine structure constant. Quantum phenomenon emerges as natural
consequence of non-differentiability of space-time. The fractal
space-time introduces a complex velocity field which has
contribution to the vacuum potential energy. It is interesting to
note that the origin of changing velocity of light has
contribution from the nature of space time. Zhang [16] has
suggested that the electric charge as a form of imaginary energy
where an electric charge is supposed to be a pack of complex
energy. The real part is supposed  to be proportional to the mass
whereas the imaginary part represent the electric charge. Electric
charge is an important component of fine structure constant. It
would be interesting to investigate how complex electric charge
affects the fine structure and if there is any connection between
the complex electric charge and complex velocity field originated
from the non-differential space-time structure. We will
investigate this problem in our future work.
\vskip 0.001in

{\bf Acknowledgement}

Authors are thankful to University Grants Commission, Govt. of INDIA for financial support.

\newpage

{\bf REFERENCES}

\vskip .1in

\noindent [1]. J. K. Webb, V. V. Flambaum, C. W. Churchill, M. J.
Drinkwater, J. D. Barrow (1999) Phys. Rev. Lett. 82(5)
884; J. K. Webb, M. T. Murphy, V. V. Flambaum, V. A. Dzuba,
J. D. Barrow, C. W. Churchill, J. X. Prochaska, A. M. Welfe (2001)
Phys. Rev. Lett. 87(9)091301; M. T. Murphy,
J. K. Webb, V. V. Flambaum, V. A. Dzuba, C. W. Churchill, J. X.
Prochaska, J. D. Barrow, A. M. Wolfe(2001) Mon. Not. Roy.
Astron. Soc. 327 1208; M. T. Murphy, J. K. Webb, V.
V. Flambaum(2003) Mon. Not. Roy. Astron. Soc.345
609; J. A. King, J. K. Webb, M. T. Murphy, V. V. Flambaum, R. F.
Carswell, M. B. Bainbridge, M. R. Wilczynska, F. E. Koch(2012)
Mon. Not. Roy. Astron. Soc. 000 1-38.

\noindent [2]. P. A. M. Dirac (1937) Nature 139 323.

\noindent [3]. J. D. Bakenstei(1982) Phys. Rev. D 25(6) (1982)1527.

\noindent [4]. M. S. Barman(2009) Rev. Mex. Astron. Astrofis.
45,139-142.

\noindent [5]. I. Haranas (2003) Rom. Astron. J. 13(1),25-30.

\noindent [6]. M. E. Tobar (2005) arXiv:hep-ph/0306230v5
.

\noindent [7]. N. Leefer, C. T. M. Weber, A. Cingoz, T. R.
Torgerson, D. Budker(2013) Phys.
Rev. Lett. 111(6) 060801.

\noindent [8]. A. F. Ranada,(2003) Euro.
Phys. Lett. 61(2) 174.

\noindent [9]. A. Bhattacharya, R. Saha and B. Chakrabarti,(2012)
Eur. Phys. J. Plus 127, 57.

\noindent [10]. L. Tedesco,(2011) Int. J. Math. Math. Sc. article ID
543894(12 pages).

\noindent [11]. J. W. Moffat,(2001)
arXiv:astro-ph/0109350v2.

\noindent [12]. L. Nottale, "Scale relativity and Fractal
Space-Time", Imperial College Press, London (2011).

\noindent [13]. Z. Chang, S. Wang, X. Li, (2012) Euro. Phys. J. C
 72, 1838.

\noindent [14].  A. Timofte, I. C. Botez, D. Scurtu, M. Agop,(2011)
Acta Phys. Pol. A  119, 304.

\noindent [15]. M. S. El Naschie,(2009) Chaos, Sol. Frac. 41, 2635.

\noindent [16]. T. Zhang,(2008) Prog. Phys. 2, 79.

\end{document}